\documentclass[12pt]{article}
\usepackage[sectionbib]{natbib}
\usepackage[]{hyperref}  
\usepackage{bm,float}
\usepackage{graphicx}
\usepackage{algorithm}
\usepackage[noend]{algpseudocode}
\usepackage[figuresright]{rotating}
\usepackage{authblk}

\title{Statistical Challenges in Modeling Big Brain Signals}
\author[1]{Zhaoxia Yu}
\author[1]{Dustin Pluta}
\author[1]{Tong Shen}
\author[2]{Chuansheng Chen}
\author[3]{Gui Xue}
\author[1, 4]{Hernando Ombao}
\affil[1]{Department of Statistics, University of California, Irvine, USA}
\affil[2]{Department of Psychology and Social Behavior, University of California, Irvine, USA}
\affil[3]{State Key Laboratory of Cognitive Neuroscience and Learning and IDG/McGovern Institute for Brain Research, Center for Collaboration and Innovation in Brain and Learning Sciences, Beijing Normal University, Beijing, PR China}
\affil[4]{Biostatistics Group, King Abdullah University of Science and Technology, Saudi\\ Email: hernando.ombao@kaust.edu.sa}

\date{\today}

\begin{document}
\maketitle 

\begin{abstract}

Brain signal data are inherently big: massive in amount, complex in 
structure, and high in dimensions. These characteristics impose great 
challenges for statistical inference and learning. Here we review several key challenges, discuss possible solutions, and highlight future research directions.

\end{abstract}

\section{Introduction}




The brain is one of the most complex organs of human body. 
Various recording techniques have been developed to get a glimpse 
of brain activity and to gain a deeper understanding of the neural 
mechanisms both during rest and in response to various stimuli. One 
of the most commonly used modalities is functional magnetic resonance 
imaging (fMRI) which measures changes in the blood oxygen level in 
localized brain regions (\cite{lindquist2008statistical}). Another important 
modality is electroencephalogram (EEG) recordings, which measures the collective electrical 
activity of populations of neurons (\cite{Ombaobook}). While both fMRI and 
EEG measure brain function over time, strucutral modalities provide additional information 
about the relationship of brain regions, for example, diffusion tensor imaging (DTI) 
maps the structure and orientation of the brain's white matter fiber tracks through the diffusion of 
water molecules across the entire brain volume (\cite{basser1994estimation}). 

Brain data recorded by these different techniques bring exciting new research opportunities, but statistical inference and analysis of these data poses immense challenges due to the sheer size of data and the complicated biological characteristics under study. For example, fMRI data consists of time series recordings over a three-dimensional image with over $100K$ voxels and sampled every $1-2$ seconds. Moreover, a typical scalp EEG is recorded across $256$ channels sampled at the rate of $1000$ observations per second.  These two brain modalities capture different aspects of brain activity (electrical by EEG and hemodynamic by fMRI) and possess different spatio-temporal resolutions (EEG has excellent temporal resolution while fMRI offers high spatial specificity). Nevertheless, they share several common features: the amount of information is massive (current studies routinely generate terabytes worth of raw data), the signal-to-noise ratio is low and the structure is complicated. In particular, studying interactions between brain regions is challenging because of the multi-scale nature of the data. One has to account for both local interactions (within voxels in a region) and global (between regions). In addition, in imaging genetics, we want to study how genetic variants are associated with brain function. Since genetic data are also high-dimensional, imaging genetics faces even more statistical and computational challenges.

\section{Brain Connectivity and Imaging Genetics}
\subsection{Brain Connectivity}

Localized brain activations play a critical role in many human brain functions (e.g., visual, motor). However, the execution of higher level cognitive 
processing (e.g., in memory retrieval, decision making) requires the
interaction and transfer of information between many of these localized 
regions. Recently, there has been great interest in studying brain connectivity, 
spurred by emerging evidence that 
connectivity may provide greater insight into a number of mental disorders, the brain-behavior relationship, and variations in cognitive performance across individuals (\cite{woodward2015resting}). 

\begin{figure}[H]
\includegraphics[scale=0.2]{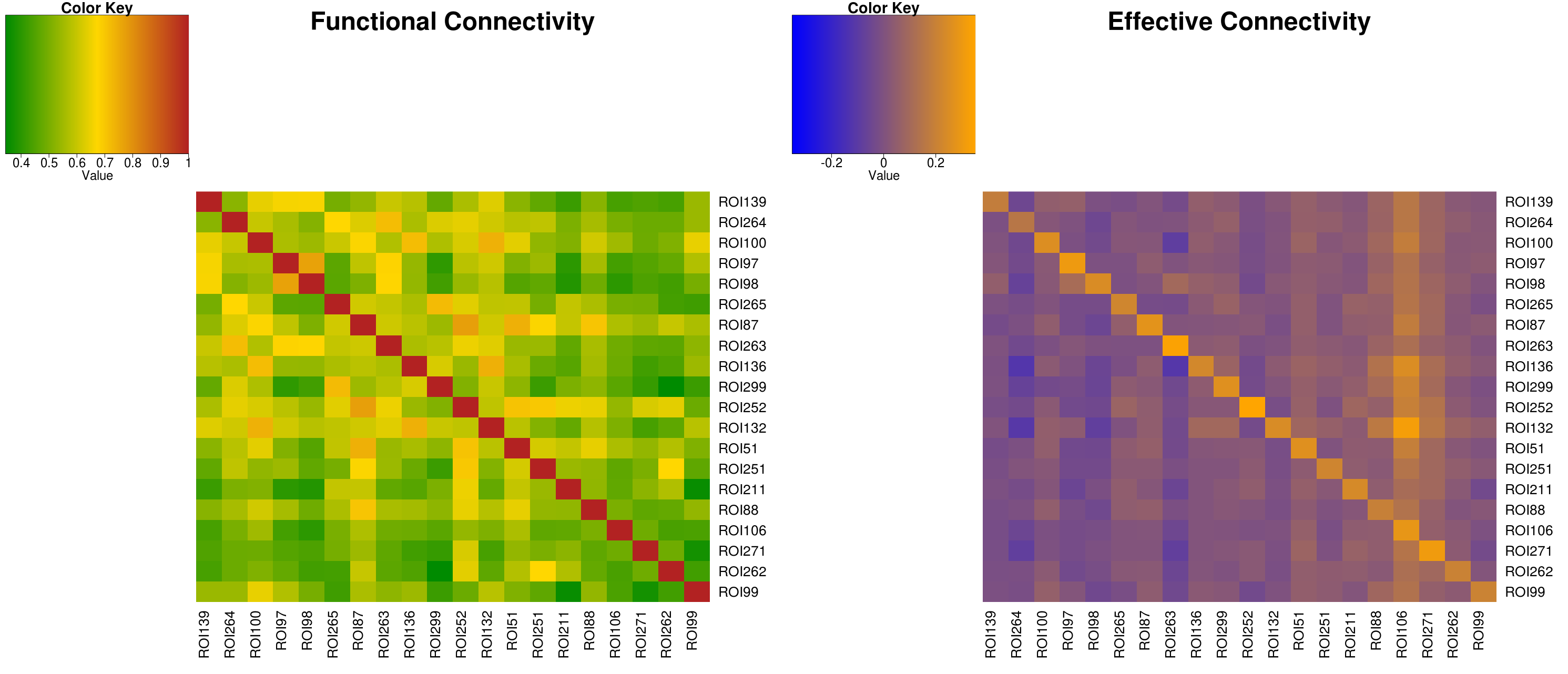}
	\caption{Functional vs. effective connectivity}
	\label{fig:connectivity}
\end{figure}

While the general idea of connectivity between two brain regions seems natural, the staggering complexity of the brain requires numerous definitions and measurement methods to highlight different aspects of brain connectivity.  There are three general concepts of brain ``connectivity" that have been of interest, namely structural, functional, and effective connectivities. Structural connectivity refers to the physical connections between brain regions and is commonly measured using DTI. Functional connectivity is a symmetric and undirected measure of concordant activation between brain regions. Some common measures of functional connectivity are cross-correlation, cross-coherence, partial-correlation, and partial coherence (see \cite{OVB,Sun,FiecasPcoh}). By contrast, effective connectivity is an asymmetric measure of how past activity in one region may influence the future activity of another region. Effective connectivity is closely related to Granger causality (as opposed to physiological causality which requires a carefully designed experiment) and is often assessed using the vector autoregressive (VAR) model (
\cite{Gorrostieta2013}, \cite{ZheYu}, \cite{BayesianVAR}). To illustrate the difference between these measures, we estimated both functional and effective connectivity using a sample of 209 subjects whose fMRI data were acquired during a cups task experiment (\cite{xue2008functional}). We found that one ROI in the right prefrontal cortex has relatively weak functional connectivity with the other ROIs. However, in terms of effective connectivity, 
it is a relatively strong {\it sender} to the other ROIs, but is a weak receiver (see Figure~\ref{fig:connectivity}).


There is an increasing body of empirical evidence indicating the importance of dynamic 
characteristics of brain connectivity for understanding neurophysiology. Interesting features of dynamic connectivity can be exhibited across widely varying time scales, ranging from 
milliseconds (in electrophysiological studies) to seconds (in fMRI studies) 
and even over years in extended longitudinal studies, 
such as studies of infant development and Alzheimer's disease progression.  See \cite{hutchison2013dynamic}
Adding this extra level of complexity further increases the difficulty 
of modelling brain connectivity, and has motivated the development of new statistical approaches.  

A variety of methods have been proposed to estimate 
dynamic connectivity, including sliding window methods (e.g., \cite{chang2010time})
change-point 
detection via VAR models (\cite{KMO}) and hidden Markov switching 
VAR models (\cite{samdin2017unified}). In a recent work,  \cite{ting2017estimating} proposed a method for high dimensional by finding low-dimensional representations via principal components 
analysis. The proposed model, f-SVAR, is a factor-based vector autoregressive model 
with Markov-switching between states. By estimating latent brain states that recur throughout the experiment, this method is able to capture connectivity patterns associated with those distinct states. 

\subsection{Imaging Genetics}
Imaging Genetics is an emerging research on determining genetic basis for brain 
anatomical structure, brain physiological activity during rest and stimulation 
and behavior. 
This new area is a potentially important tool in early diagnosis, personalized medicine, and 
the treatment of neurological disorders 
(\cite{stingo2013integrative}). Indeed, numerous work has been recently published to assess the genetic components of structural and functional imaging data. In particular, 
\cite{ge2016multidimensional} demonstrate that several characteristics of brain structure are genetically heritable, such as brain volume and neuroanatomical shape. As we move from structural imaging towards functional imaging, the 
magnitude of data complexity sharply increases. This is 
especially true for brain connectivity, which examines how genetics 
may explain variation in between-regions interactions. Recent work suggests that brain connectivity has the qualities of a fingerprint, i.e., everyone 
one has an own unique profile regardless at which condition the connectivity 
is measured (\cite{finn2015functional}). Like other characteristics of the 
brain, and probably more so due to its ``fingerprinting" property, 
connectivity is expected to be heritable. Indeed, a recent study based 
on families with attention-deficit/hyperactivity disorder (ADHD) suggests significant genetic heritability for the default mode, cognitive control, and ventral attention networks (\cite{sudre2017estimating}). 

\section{Statistical Challenges and Strategies}
It is widely accepted that statistical inference for neuroimaging data 
face tremendous challenges. As noted, imaging data are inherently 
multi-modal and high dimensional. Moreover, they suffer from low signal-to-noise ratios and small effect sizes from individual genetic factors and brain regions. We review several statistical strategies. 

\subsection{Multiple Testing}
Multiple testing is a common and serious issue in analyzing brain imaging 
data as tests of activation are performed on $100$K voxels or thousands 
of regions of interest  (ROIs) in fMRI and hundreds of channels in EEG. 
In addition, there are tests on connectivity across all pairs of voxels, pairs of 
ROIs and pairs of channels (\cite{nichols2012multiple}). The burden of 
multiple testing is further amplified in imaging genetics, where the curse 
of dimensionality comes from both the brain signal data and the genetic 
variables 
. In the massive univariate approach, 
each pair of voxel and single nucleotide polymorphism (SNP) will be 
tested, leading to trillions of tests (\cite{stein2010voxelwise}), 
Given this huge number of tests, it is then necessary to control the expected number 
of false positives. One method is to control 
the familywise error rate (FWER), i.e., the probability of making at least 
one false positive, at a pre-specified value. A simple method is 
Bonferroni correction where the p-value cutoff is defined as the targeting 
FWER divided by the number of tests. Although Bonferroni correction successfully controls FWER, it tends to be overly conservative 
and thus can potentially fail to detect important features in the data. Instead, in many 
situations, it makes sense to incur the cost of having a small proportion 
of false positive in return for more true positives. Consequently, it is reasonable to control the false discovery rate (FDR), which is defined as the expected proportion of false positives among the ones that have been declared to be significant (\cite{benjamini1995controlling}). 
However, for imaging analysis the issue is further complicated by the spatial dependence of the data, which can decrease overall power if ignored. To address this, resampling methods such as permutation-based evaluation can be used (\cite{nichols2002nonparametric, 
razpermutation}), which have the advantage of preserving the correlation structure within a modality. On the other hand, permutations are computationally expensive, and can be impractical when the computational cost of each permutation is substantial. 

\subsection{Dimension Reduction, Regularized Analysis, and Low-Rank Approximation}
Due to the numerous challenges posed by high-dimensional data, dimension reduction is often necessary for efficient statistical analysis of imaging and genetics data.  Some commonly used data drive methods 
include principal component analysis (PCA), high order singular 
value decomposition (the high dimension extension of PCA), and independent component analysis (ICA)(\cite{liu2009combining}). These methods, although effective in reducing dimensions for the genetic or imaging modality, are often suboptimal for prediction and testing, as the extracted leading components do not necessarily capture the strongest associations between genetic and imaging modalities.  Simultaneous dimension reduction methods, such as partial least squares (PLS), canonical correlation analysis (CCA), reduced rank regression (RRR), and parallel ICA 
\cite{ahn2015sparse}
on the other hand, perform dimension reduction jointly on responses and covariates by minimizing a specific cost function. 

Recently, tensor regression has been of increasing interest for analyzing multi-way and high-dimensional data such as brain imaging data. By analyzing tensors, i.e., multi-array data, tensor regression can directly incorporate the inherent spatio-temporal structure of imaging data. At the same time, it achieves parsimony by imposing low-rank decompositions (such as PARAFAC) and sparsity assumptions (such as LASSO) on the regression coefficients. By utilizing these reduction methods, which depend on the response data as well as the covariates, tensor regression methods may lead to more accurate parameter estimation and outcome prediction (\cite{lu2017bayesian, 
zhou2013tensor}) 
An additional advantage of these approaches is that individual coefficients between covariates and responses can be estimated, which makes the results more interpretable than those from traditional dimension reduction methods. Indeed, \cite{zhou2013tensor} identified brain regions that might be responsible for ADHD by applying tensor regression on 3D MRI data. It is likely that extensions of these methods, such as adding a time dimension with hidden states, will improve our understanding of dynamic connectivity and how it is connected to behavioral traits, cognitive outcomes, and genetic factors.

\subsection{Global Tests}
The random effect model, also known as variance component model in this setting, has been proposed to jointly model a large number of genetic variants 
(\cite{ge2016multidimensional}). In this approach, the effect from an individual genetic variant is treated as a random variable, rather than a fixed parameter. As a result, the number of variants that can be jointly analyzed is not limited by sample size. It is also noticed that this method is closely related to kernel machine methods, which can naturally model non-additive effects 
(\cite{ge2012increasing}). These approaches are often used to analyze sets of genetic variants, such as the variants in a gene or pathway, or in estimating the genetic heritability of a phenotype. For hypothesis testing, score tests are often used because they can be calculated rapidly. 

We recently discovered that these score tests have a unified form (\cite{dustin2017mantel}), which we refer to as Mantel's statistic (\cite{mantel1967detection}). Essentially, this approach examines whether there is a relationship between two sets of variables by testing whether the pairwise similarity/distance measurements between the two sets of variables are correlated. This is a very general framework that includes many other proposed approaches as special cases, such as multivariate distance matrix regression, pseudo F-tests, and kernel-based tests (\cite{shehzad2014multivariate, 
xu2017adaptive}). Moreover, this method is not limited in application to quantitative and matrix-shaped data, but can be implemented as long as there is a suitable definition of similarity between two subjects. In theory, one can also model interactions between all pairs of genes/SNPs, high-order interactions, and even complicated and unknown functions of genetic variants, through an appropriate choice of similarity measure.  This flexibility makes the framework well-suited for testing the association of two sets of high-dimensional, biologically complex features. 

\subsection{Data Integration}


Similar to dimension reduction methods, data integration methods also aim to improve efficiency and power. However, different from dimension reduction, data integration achieves better power by aggregating weak signals from multiple sources. For example, one can conduct meta-analysis by combining data collected from different centers (\cite{wager2007meta, salimi2009meta}). Another direction, particularly useful for neuroimaging studies, is to integrate information from different imaging modalities, as different modalities often measure complementary characteristics of brain function. For example, fMRI has high spatial but low temporal resolution; on the contrary, EEG has low spatial but high temporal resolution. Multi-modal data can be acquired simultaneously or separately. While simultaneous acquisition sounds appealing, it faces many technical issues. For example, when fMRI and EEG data are recorded during the same experiment, EEG electrodes can distort the magnetic field, making the fMRI data less accurate (\cite{uludaug2014general}). As a result, we here focus on combining information from separately (non-contemporaneously) recorded data. Previous work in this direction has fused fMRI and DTI data in estimating brain connectivity (
\cite{
, xue2015multimodal, kang2017bayesianfusing}), and using structural connectivity from DTI to construct informative priors when estimating resting state functional connectivity from fMRI (\cite{kang2017bayesianfusing,BayesianVAR}).


\section{Discussion}
Brain signals have several unique characteristics. Its low signal-to-noise ratio, 
the difficulty to select truly associated features in high-dimensional data, and the small individual effect sizes (such as genetic effects) all contribute to the difficulties and low statistical power. In particular, as we are expanding analysis from individual ROIs/voxels to connectivity and from connectivity to dynamic connectivity, there is an urgent need for computationally and statistically efficient methods. 

Given the various issues in brain signal and other related data, methods for data integration would be valuable. Indeed, methodological development on integrating different brain signal data is a fast evolving area. A practical and useful strategy to aggregate weak signal is to conduct global tests between two complicate and high-dimensional domains. Taking imaging genetics as an example, given the statistical chellenges for realistic sample sizes, a logical first step is to examine whether there is an ``overall" association between the brain connectivity modality and the genetic modality. Considering another example, for brain connectivity separately estimated from fMRI and EEG, it is of interest to know the concordance of these measures.  Measuring the consistency of connectivity measures across modalities can lead to improved understanding of brain connectivity in general, and inform future statistical analyses.  For instance, our recent work has applied the Mantel test to show that brain connectivity estimated using fMRI is consistent with that using EEG (\cite{dustin2017mantel}), suggesting that it may be possible to develop more powerful models of brain connectivity by combining fMRI and EEG data. With continued improvements in imaging and genetic sequencing technologies, and the availility of relevant publicly available data banks, it is clear that there will be a continued need for further developing a variety of flexible, robust statistical methods for multi-modal data.

\bibliographystyle{apalike}
\bibliography{ref}

\end{document}